\documentclass[compsoc,conference,a4paper,10pt]{IEEEtran}
\IEEEoverridecommandlockouts

\usepackage[all,defaultlines=3]{nowidow} 
\usepackage{enumitem}
\usepackage{graphicx}
\usepackage{xspace}
\usepackage{listings}
\usepackage[frozencache=true,cachedir=_minted-main]{minted}
\usepackage[nocompress]{cite}
\usepackage{caption}
\usepackage{multicol}
\usepackage{sistyle}
\usepackage{parcolumns}
\usepackage{float}
\usepackage{algorithm}
\usepackage{algorithmic}
\usepackage{adjustbox}
\usepackage{wrapfig}
\usepackage{xcolor}
\usepackage{tikz}
\usetikzlibrary{positioning,arrows.meta}
\usepackage{pifont}
\usepackage{multirow}
\SIthousandsep{,}

\usepackage[colorlinks=true,urlcolor=blue,citecolor=blue,linkcolor=blue]{hyperref}

\setminted{fontsize=\footnotesize}

\renewcommand{\footnoterule}{\kern-3pt\hrule width \columnwidth\kern 2.6pt}

\newcommand{\system}{{\textsc{Scribe}}\xspace} 

\newcommand{\circled}[1]{\texorpdfstring{
  \tikz[baseline=(char.base)]{
    \node[shape=circle,fill=black,inner sep=0.8pt,text=white] (char) {#1};}%
  }{#1}}

\begin{document}

\title{SCRIBE: Practical Static Binary Patching via Binary-Aware Recompilation of Decompiled Code}

\author{\IEEEauthorblockN{Han Dai}
\IEEEauthorblockA{
\textit{Purdue University}\\
dai124@purdue.edu}
\and
\IEEEauthorblockN{Soumyakant Priyadarshan}
\IEEEauthorblockA{
\textit{Purdue University}\\
priyadar@purdue.edu}
\and
\IEEEauthorblockN{Abdullah Imran}
\IEEEauthorblockA{
\textit{Purdue University}\\
imran8@purdue.edu}
\and
\IEEEauthorblockN{Ruoyu Wang}
\IEEEauthorblockA{
\textit{Arizona State University}\\
fishw@asu.edu}
\and
\IEEEauthorblockN{Antonio Bianchi}
\IEEEauthorblockA{
\textit{Purdue University}\\
antoniob@purdue.edu}
}

\maketitle

\makeatletter
\long\def\IEEEcopyrightfntext#1{\begingroup
  \renewcommand\@IEEEcompsocmakefnmark{}%
  \xdef\@thefnmark{}%
  \@footnotetext{#1}%
\endgroup}
\makeatother
\IEEEcopyrightfntext{\copyright~2026 IEEE. Personal use of this material is permitted. Permission from IEEE must be obtained for all other uses, in any current or future media, including reprinting/republishing this material for advertising or promotional purposes, creating new collective works, for resale or redistribution to servers or lists, or reuse of any copyrighted component of this work in other works.}

\begin{abstract}
When source code or the original toolchain is unavailable, patching binaries is difficult because it requires editing low-level assembly code directly.
As an alternative, one can decompile the binary, apply the patch at the source level, and then recompile the modified code.
However, as this paper demonstrates, this workflow is hindered by pervasive syntactic and semantic inaccuracies in the output of modern decompilers, many of which prior work has overlooked.

To address these challenges, we present \system, a patching framework that handles syntactic and semantic issues in decompiled code, improving both recompilation success and correctness.
\system's novel ``binary-aware'' recompilation approach repairs semantic inaccuracies in decompiler output by leveraging information extracted directly from the original binary.
In our evaluation, \system resolved approximately 81\% of previously incorrect functions produced by the Hex-Rays decompiler, demonstrating the effectiveness of its approach.
Moreover, we show that, using \system, it is possible to patch 13 of 14 real-world CVEs without access to the original source code and without performing any manual binary editing.
To further validate our findings, we conducted a user study with 18 participants. Using \system, participants achieved 100\% patching success, compared to 3.7\% without it.
Finally, we asked three large language models to generate source-level patches via \system; all three achieved 100\% success when using the framework, demonstrating its potential to enable fully automated patching.
Overall, these results indicate that \system makes source-level patching of binaries accessible and reliable, even without access to the original source.
\end{abstract}



\section{Introduction}
\label{sec:intro}

Rapid application of software patches is essential for maintaining the security and reliability of modern systems. In practice, however, patch deployment is often impeded by the absence of source code or compatible build toolchains. Many closed-source applications, particularly abandoned or no longer vendor-supported ones, offer no access to the original development environments. Commercial off-the-shelf (COTS) software frequently includes statically linked or bundled third-party libraries with known vulnerabilities~\cite{b2sfinder}. Even when patches for these libraries are available, updating them can be infeasible: the application may depend on a specific version of the library, with a build environment (e.g., OS, compiler, dependencies) that is outdated, undocumented, or irreproducible~\cite{osspatcher}. In some cases, libraries rely on proprietary components that are inaccessible or legally restricted. Additionally, applications may incorporate heavily customized versions of libraries, where upstream patches do not apply cleanly and require nontrivial manual adaptation. These challenges make source-level patching impractical and highlight the need for binary-level solutions.

As a result, vulnerabilities often remain unfixed or \emph{orphaned}, silently accumulating in widely deployed systems. Indeed, recent studies~\cite{reid2022extent} reveal that such orphan vulnerabilities are pervasive, even in mature and well-maintained open-source software (OSS) ecosystems. Without patching these orphan vulnerabilities promptly, users of affected software remain vulnerable to security threats indefinitely. While researchers have made significant progress in identifying orphan vulnerabilities in binary code~\cite{pewny2015cross,eschweiler2016discovre,luo2023vulhawk}, practical and reliable methods for \emph{fixing} such vulnerabilities remain elusive, even when source-level patches are available.
Binary patching~\cite{dyninst,e9patch} offers a potential solution but typically requires deep reverse-engineering expertise, including familiarity with low-level assembly, the target architecture, and the instruction set, because it requires identifying the assembly instructions to patch, understanding how to modify them, and manually writing any additional instructions the patch may require.
In fact, as our user study shows, patching at the binary level remains a time-consuming and error-prone process, even for seasoned practitioners.

Existing work, such as OSSPatcher~\cite{osspatcher}, explores applying source-level patches to binaries but assumes access to the original source code and matching toolchains—an assumption that rarely holds for legacy or closed-source software. Other tools~\cite{binrec,wytiwyg,mcsema,psi,safer,egalito,stir,reins} focus on lifting binaries to assembly or intermediate representations (IR) to enable recompilation, but do not support direct source-level patching.

To address these challenges, we explore a novel approach: enabling the application of source-level patches directly to binary executables.
This approach not only lowers the barrier for developers not familiar with binary analysis, but also opens the door to leveraging automated program repair (APR) techniques~\cite{apr-abhik,monperrus2018living,neural-repair,vulrepair}, which are designed to operate at the source level.



To enable our goal of applying source-level patches to binaries, our proposed methodology requires decompiling individual functions, applying source-code modifications, and seamlessly reintegrating the changes into the original binary.
In turn, this capability requires being able to recompile decompiled source code.
However, recompiling decompiled source code is challenging because modern decompilers~\cite{ida,ghidra} are far from producing ``perfect'' decompiled code, and they generally focus on producing readable code for manual analysis rather than recompilable code.
Consequently, their output often suffers from syntactic and semantic inaccuracies. Syntactic issues prevent compilation entirely, while semantic issues are more subtle. In this case, the code compiles, but exhibits incorrect runtime behavior due to misidentified types, structs, or disrupted memory layouts. For example, if the decompiler fails to identify a struct type and instead identifies the members as individual variables, the decompiled code will still compile, but the compiler might reorder the layout of these variables, leading to incorrect memory access.

To address these issues, we propose \system\footnote{\system{} is a recursive acronym for ``\textsc{Scribe} Can Recompile and Inject Binary Edits.''}, a recompilation framework that uses \emph{binary-aware recompilation}: instead of attempting to fix the decompilers themselves, \system{} strategically applies targeted remedies throughout the recompilation process. It focuses on minimal invasiveness, only recompiling and retrofitting the functions required for deploying a patch, reducing complexity and risk. Crucially, \system{} preserves low-level properties (such as stack layouts and variable arrangements) necessary for seamless integration with the original binary.

We demonstrate \system{}'s effectiveness in reducing the number of incorrectly recompiled functions and, through a controlled user study, show that it enables developers without reverse-engineering expertise to reliably patch security vulnerabilities in closed-source binaries. Our empirical study demonstrates that \system{} reconstructs functional code for approximately 81\% of previously unworkable functions produced by Hex-Rays across various optimization levels.

Building on this recompilation capability, we then show that \system{} enables the successful patching of 13 out of 14 real-world CVEs in GNU Coreutils and Binutils, all without access to the original source code or build environments.
Most importantly, our user study highlights \system{}'s practical impact.
In this study, we show that 18 participants with diverse backgrounds and skill levels were able to patch every vulnerability presented to them when using \system{}.
In contrast, without \system{}, the same participants achieved only a 3.7\% success rate, demonstrating that \system{} drastically improves correctness and efficiency in patching binaries. Finally, an evaluation conducted with three state-of-the-art LLMs (GPT-5, Claude 4.5 Sonnet, and Gemini 2.5 Pro) produced similar outcomes, with 100\% success when using \system{} and 0\% without it.

Overall, these results show that \system{} substantially improves recompilation reliability and enables practical and accurate binary patching, making the process accessible even to developers who have no reverse-engineering experience.

In summary, this paper makes the following contributions:
\begin{itemize}[leftmargin=*]
    \item We systematically study the challenges of recompiling decompiled code,
      and we unveil previously unknown issues that prevent correct recompilation.
    \item We propose \system, which uses a novel
      \emph{binary-aware recompilation} technique to fix syntactic and semantic issues and significantly improve recompilation of decompiled source code.
    \item We evaluate \system with Hex-Rays~\cite{ida} and Ghidra~\cite{ghidra}
      decompilers on the latest versions of GNU Coreutils and CGC binaries at
      4 optimization levels, demonstrating substantial improvements in recompilation correctness
      for both decompilers across diverse programs and optimization settings.
    \item We demonstrate \system's real-world applicability by successfully
      patching reproducible CVEs in GNU Coreutils and GNU Binutils
      without access to original source code or build environments.
    \item We validate \system's practical usability through a controlled
      user study with 18 participants of varying expertise and automated evaluations with
      3 state-of-the-art LLMs, demonstrating dramatic improvements over traditional approaches
      for both human developers and LLMs.
\end{itemize}

To foster reproducibility and support future research, we make all source code, datasets, and accompanying documentation publicly available at \url{https://github.com/purseclab/scribe}.


\section{Background and Motivation}

\begin{figure}[t]
\centering
\begin{tikzpicture}[
    box/.style={rectangle, draw, thick, minimum width=2.1cm, minimum height=0.6cm, align=center, font=\scriptsize},
    smallbox/.style={rectangle, draw, thick, minimum width=2.1cm, minimum height=0.5cm, align=center, font=\scriptsize},
    interventionbox/.style={rectangle, draw, thick, fill=gray!20, minimum width=2.1cm, minimum height=0.5cm, align=center, font=\scriptsize},
    arrow/.style={->,>=stealth, thick},
    dasharrow/.style={->,>=stealth, thick, dashed, gray},
    title/.style={font=\small\bfseries},
    note/.style={font=\scriptsize, align=center, text width=2.1cm}
]

\node[title] (t1) at (0,0) {Direct Rewriting};
\node[box, below=0.3cm of t1] (b1) {Binary};
\node[interventionbox, below=0.25cm of b1] (m1) {Modify\\Assembly};
\node[smallbox, below=0.25cm of m1] (p1) {Patched\\Binary};
\node[note, below=0.1cm of p1] (scope1) {(Localized)};
\draw[arrow] (b1) -- (m1);
\draw[arrow] (m1) -- (p1);

\node[title] (t2) at (2.7,0) {IR/ASM Lifting};
\node[box, below=0.3cm of t2] (b2) {Binary};
\node[smallbox, below=0.25cm of b2] (l2) {Lift to\\IR/Assembly};
\node[interventionbox, below=0.25cm of l2] (m2) {Modify\\IR/Assembly};
\node[smallbox, below=0.25cm of m2] (r2) {Recompile/\\Reassemble};
\node[smallbox, below=0.25cm of r2] (p2) {Patched\\Binary};
\node[note, below=0.1cm of p2] (scope2) {(Whole-program)};
\draw[arrow] (b2) -- (l2);
\draw[arrow] (l2) -- (m2);
\draw[arrow] (m2) -- (r2);
\draw[arrow] (r2) -- (p2);

\node[title] (t3) at (5.4,0) {SCRIBE};
\node[box, below=0.3cm of t3] (b3) {Binary};
\node[smallbox, below=0.25cm of b3] (d3) {Decompile\\Target Funcs};
\node[interventionbox, below=0.25cm of d3] (pa3) {Patch C\\Code};
\node[smallbox, below=0.25cm of pa3] (r3) {Binary-aware\\Recompile};
\node[smallbox, below=0.25cm of r3] (p3) {Patched\\Binary};
\node[note, below=0.1cm of p3] (scope3) {(Per-function)};

\draw[arrow] (b3) -- (d3);
\draw[arrow] (d3) -- (pa3);
\draw[arrow] (pa3) -- (r3);
\draw[arrow] (r3) -- (p3);

\draw[dasharrow] (b3.east) -- ++(0.4,0) |- (r3.east);

\end{tikzpicture}
\caption{Comparison of binary patching approaches. Human intervention steps are highlighted in gray. SCRIBE uses binary-aware recompilation: metadata from the original binary, shown by the dashed arrow, guides recompilation to preserve low-level properties.}
\label{fig:patching_approaches}
\end{figure}
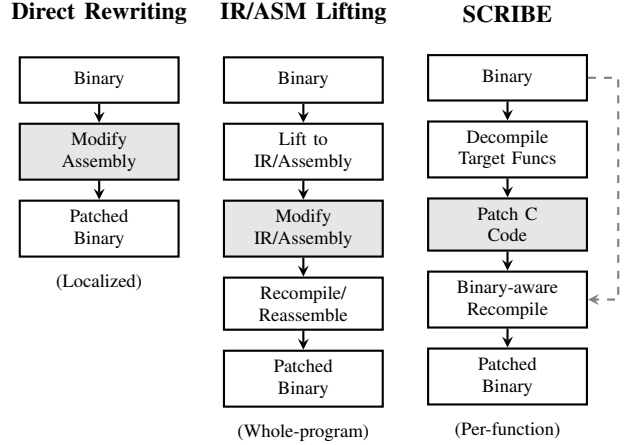

The primary goal of this work is to enable applying source-level patches directly to binaries.
Patching binaries without access to source code is essential for maintaining legacy systems, closed-source software, and IoT devices where vendors may not provide timely security updates.
To understand why decompilation-based patching is necessary, we first examine existing binary patching approaches and their limitations.
Figure~\ref{fig:patching_approaches} illustrates three paradigms, each with distinct trade-offs.

\textbf{Direct Binary Rewriting.}
Tools like Patcherex2~\cite{patcherex2}, E9Patch~\cite{e9patch}, and Dyninst~\cite{dyninst} enable patching by directly modifying binaries at the assembly or instruction level.
While these tools can handle space management through techniques like adding new segments, using trampolines, or finding padding, applying patches requires operating at the assembly level with deep understanding of low-level details such as calling conventions, register usage, and memory layouts.
This makes complex semantic patches that are naturally expressed by using source code difficult to implement and maintain.

\textbf{Lifting and Reassembly.}
An alternative approach lifts entire binaries to intermediate representations or reassemblable assembly for modification and recompilation.
Tools like BinRec~\cite{binrec} and McSema~\cite{mcsema} lift to LLVM IR, while Ramblr~\cite{ramblr} and Egalito~\cite{egalito} work with reassembled assembly.
Since these tools recompile or reassemble the entire binary, they can freely extend code without the space constraints of direct rewriting.
However, they face scalability challenges when processing large binaries and still require developers to work at relatively low abstraction levels like LLVM IR or assembly.
More critically, applying source-level patches designed for C code to IR or assembly remains non-trivial, limiting their utility for leveraging existing patch repositories and automated repair techniques.

\textbf{Decompilation-based Patching.}
To enable source-level patching without source code access, we adopt a decompile-patch-recompile workflow.
Starting with a binary and a source-level patch, we decompile target functions into C using tools like Hex-Rays~\cite{ida} or Ghidra~\cite{ghidra}.
Developers or automated tools then apply the patch to the decompiled source.
Finally, we recompile the patched functions and link them back into the original binary.
This approach allows working with familiar source-level code and patches while operating directly on binaries.
OSSPatcher~\cite{osspatcher} explores similar ideas but assumes access to matching source code and toolchains, an assumption that rarely holds for legacy or closed-source software.

The critical challenge in decompilation-based recompilation lies in handling decompiler inaccuracies.
Binaries compiled without debug symbols lack high-level information such as types, variable names, and structure definitions, forcing decompilers to approximate the original source.
These approximations introduce both syntactic errors that prevent compilation and semantic errors that cause incorrect runtime behavior.
As our user study demonstrates in Section~\ref{sec:eval}, even experienced practitioners struggle with naive decompilation-based patching, achieving only 3.7\% success without specialized techniques.
\system addresses these challenges through \emph{binary-aware recompilation}: by extracting and preserving low-level properties from the original binary during recompilation, \system ensures binary compatibility even when the decompiled source contains structural inaccuracies.


\section{Challenges in Recompilation}
\label{sec:challenges}
We broadly categorize the challenges associated with recompiling decompiled
code into {\em Syntactic issues} and {\em Semantic issues}.  {\em Syntactic
issues} are general violations of the programming language rules regarding
variable naming conventions, function definitions, etc. These issues cause
the compiler to report errors and abort before creating an object file. {\em
Semantic issues} are more subtle and are not caught by the compiler. They make
the behavior of recompiled code diverge from the original. Additionally, for
our approach, semantic issues also include challenges involved in relinking the
recompiled function with the original binary.

We identified these challenges through systematic analysis of compilation
failures across GNU Coreutils and CGC binaries decompiled with Hex-Rays and
Ghidra. Functions often exhibited multiple overlapping issues requiring iterative
fixes. We manually categorized error types and traced root causes by using GDB
to debug both original binaries and recompiled code, comparing execution behavior
to identify decompilation errors.

\subsection{Challenge \protect\circled{1}: Syntactic issues}
\label{sec:challenge_syntactic}
A recent study performed by Liu et al.~\cite{evaluating-recompilation} showed that
only 50\% of code produced by the best-performing decompiler (Hex-Rays) is syntactically correct. 
Compilers will refuse to process code with syntactic issues, requiring these 
problems to be resolved before any recompilation attempt.
Common syntactic issues include: (i) malformed names (e.g.,
from \texttt{int[5] name} to \texttt{int name[5]}), (ii) mismatches in function prototypes between call
sites and callees (e.g., when a caller attempts to read a return value from
a function that returns void), and (iii) missing definitions of structs, enums,
types, macros, and global variables.  For instance, decompilers often generate
decompiler-specific macros to represent sets of low-level instructions or
bitwise operations. While these macros are typically straightforward for humans
to understand, they lack actual implementations that compilers can process.

\begin{listing}
\begin{minted}{c}
int blake2b_init(blake2b_state *S, size_t outlen)
{
  blake2b_param P[1];

  if ((!outlen) || (outlen > BLAKE2B_OUTBYTES))
    return -1;

  P->digest_length = (uint8_t)outlen;
  P->key_length    = 0;
  P->fanout        = 1;
  P->depth         = 1;
  store32(&P->leaf_length, 0);
  store32(&P->node_offset, 0);
  store32(&P->xof_length, 0);
  P->node_depth    = 0;
  P->inner_length  = 0;
  memset(P->reserved, 0, sizeof( P->reserved ));
  memset(P->salt,     0, sizeof( P->salt ));
  memset(P->personal, 0, sizeof( P->personal ));
  return blake2b_init_param( S, P );
}
\end{minted}
\begin{minted}{c}
int64 blake2b_init(int64 a1, unsigned int64 a2)
{
  char v4[4]; // [rsp+30h] [rbp-50h] BYREF
  char v5[4]; // [rsp+34h] [rbp-4Ch] BYREF
  char v6[4]; // [rsp+38h] [rbp-48h] BYREF
  char v7[6]; // [rsp+3Ch] [rbp-44h] BYREF
  char s[14]; // [rsp+42h] [rbp-3Eh] BYREF
  char v9[16]; // [rsp+50h] [rbp-30h] BYREF
  char v10[24]; // [rsp+60h] [rbp-20h] BYREF
  unsigned int64 v11; // [rsp+78h] [rbp-8h]

  v11 = __readfsqword(0x28u);
  if ( !a2 || a2 > 0x40 )
    return 0xFFFFFFFFLL;
  v4[0] = a2;
  v4[1] = 0;
  v4[2] = 1;
  v4[3] = 1;
  store32(v5, 0LL);
  store32(v6, 0LL);
  store32(v7, 0LL);
  v7[4] = 0;
  v7[5] = 0;
  memset(s, 0, sizeof(s));
  memset(v9, 0, sizeof(v9));
  memset(v10, 0, 0x10uLL);
  return blake2b_init_param(a1, v4);
}
\end{minted}
    \caption{Difference between the original (top) and the decompiled code (bottom) for a stack local structure}
    \label{lst:b2sum_blake2b_init}
\end{listing}
\subsection{Challenge \protect\circled{2}: Memory layout distortions} \label{memlayoutissue}
When recompiling from decompiled high-level code, the layout (order,
padding, alignment, etc.) of local variables on the stack may not match the
original binary, leading to incorrect memory accesses. Detecting the bounds of
local objects (e.g., arrays and structures) on the stack is challenging.
Furthermore, decompilers often fail to
reconstruct original struct definitions, instead treating them as a series of
individual variables. In this scenario, the decompiled code of a given
function, at the source-code level, may still be equivalent to the original
source code.
However, during recompilation, the compiler may decide to use a different struct layout than that in the original binary.
As an example, Listing~\ref{lst:b2sum_blake2b_init} shows source code and decompiled code for
the same function that initializes a struct on the stack and fills it with
values. Although the decompiled code preserves the overall logic of the function
accurately, it fails to recover the struct definition. Instead, it
represents the struct members as separate variables (\texttt{v4} to \texttt{v10} and \texttt{s}).
This discrepancy can cause significant issues when recompiling the decompiled code.
During recompilation, the compiler may arrange these variables differently in
memory compared to the original struct and ignore any padding present in
the original struct. This can result in incorrect memory access, especially when the
structure is passed as an argument to another function. 

\subsection{Challenge \protect\circled{3}: Linking recompiled functions to the original binary} \label{ptridenterr} 

When retrofitting new recompiled functions, the primary challenge is to
preserve their compatibility with the rest of the unaltered binary. First, we
cannot relocate any of the unaltered functions or data objects to make space for
the new function. Otherwise, data and code references (e.g., function calls) will
not work correctly. Second, we need to ensure that the unaltered functions can call the recompiled function. The function might also be reachable via pointers
(e.g., virtual functions). Since identifying pointers in binaries is difficult due
to the lack of symbols and type information~\cite{uroboros,ddisasm,ramblr},
preserving pointer-based function calls to the recompiled function can be
challenging. Third, the recompiled function also needs to access other functions
and data objects and may use pointers to do so. Decompilers use static
analysis to recover pointer constants, but such static analysis is prone to
errors~\cite{ramblr,ddisasm}.
\section{System Design}
\label{sec:design}
\begin{figure}
\centering
  \includegraphics[width=\linewidth]{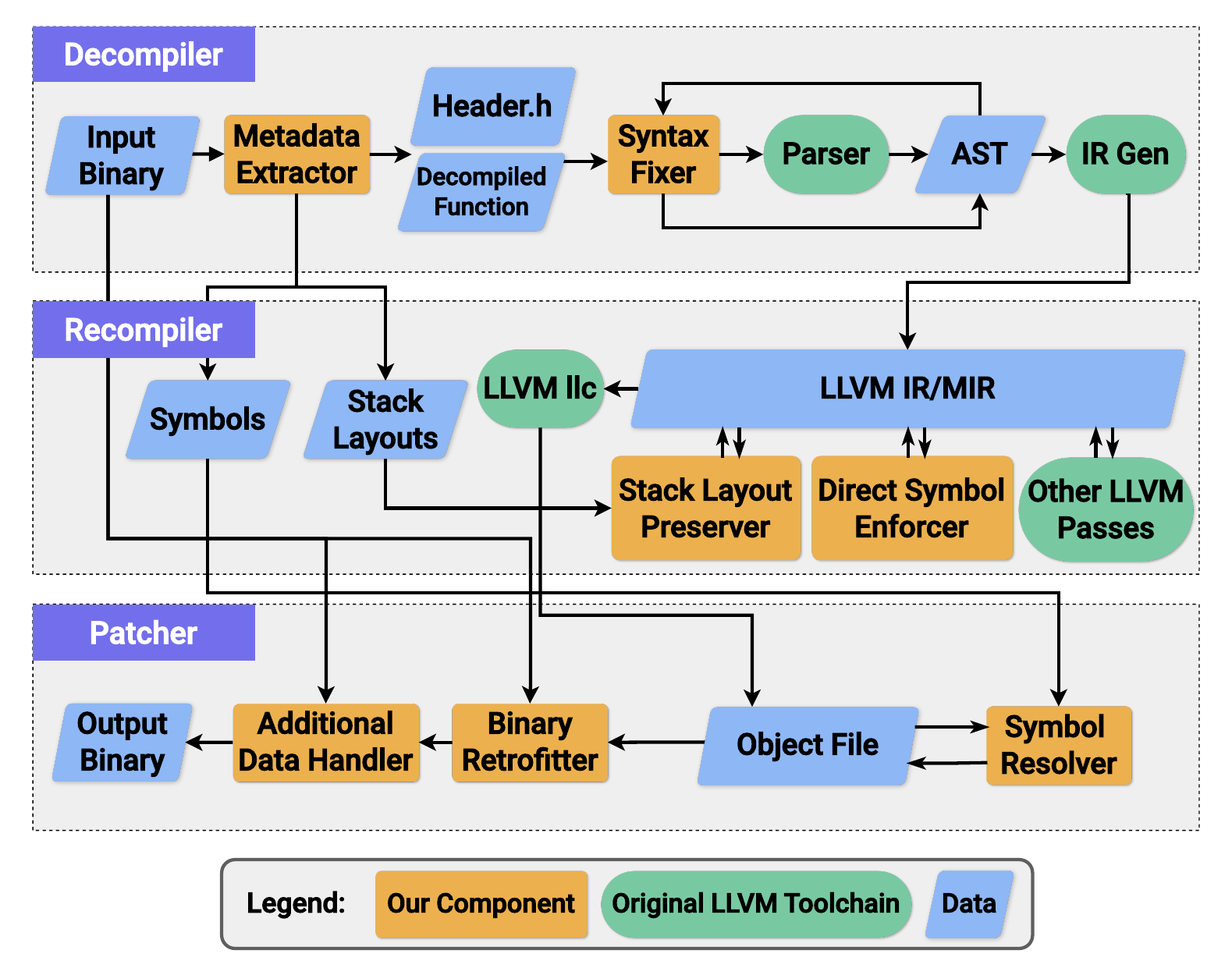}
  \caption{Overview of \system}
  \label{fig:overview}
\end{figure}

\system employs a minimally invasive approach to binary patching. Rather than recompiling entire programs, it selectively recompiles only functions needing modification, ensuring seamless interoperability between recompiled components and unmodified binary portions. This preserves critical low-level aspects including memory layouts, variable arrangements, function/data references, and control flow. \system's architecture consists of three main components as shown in Figure~\ref{fig:overview}:
\begin{itemize}
  \item \textbf{Decompiler:} Decompile specific function(s) into C code,
    apply syntactic fixes addressing Challenge~\circled{1}, and collect metadata needed by other components.
  \item \textbf{Recompiler:} Apply semantic fixes addressing Challenge~\circled{2} and compile the decompiled code into an object file.
  \item \textbf{Patcher:} Retrofit the recompiled object file into the original
    binary and ensure its compatibility with the original binary by addressing
    Challenge~\circled{3}.
\end{itemize}

\subsection{Decompiler} \label{decomp}
The decompiler component wraps state-of-the-art decompilers. The current
implementation supports Hex-Rays and Ghidra, but its modular design supports easy
extension to others. It utilizes headless-ida~\cite{headless-ida} for Hex-Rays
and pyhidra~\cite{pyhidra} for Ghidra to enable automated interaction. It contains the
following subcomponents: 

\subsubsection{Syntax Fixer} \label{syntax_fixer}
The syntax fixer addresses Challenge~\circled{1} by handling many syntax issues in the decompiler-produced code, ensuring the fixed decompiled code can be compiled.

\textbf{Correcting malformed C code.}
\system normalizes malformed names in the decompiled code. In particular, (i) it
removes non-standard symbols (., @, \$) from names and (ii) fixes incorrectly
formed array declarations (e.g., \texttt{int[10] var;} instead of \texttt{int
var[10];}). Additionally, \system removes non-standard keywords
(\texttt{\_\_noreturn}, \texttt{\_\_usercall}, \texttt{\_\_thiscall}) produced
by Hex-Rays.


\textbf{Adding definitions for undefined macros and types.}
Decompilers often generate ``generic'' custom types based only on size inference (e.g., \texttt{BYTE}, \texttt{WORD}, \texttt{DWORD}). \system ensures error-free compilation by providing definitions through a specially crafted header file that includes decompiler-specific macros. In the presence of binary-level security features, such as stack canaries, decompilers often produce non-standard C instructions (such as instructions using the \texttt{\_\_readfsqword} macro) to indicate their presence.
To enable proper recompilation of such code and generate code with analogous security features, we replace these non-standard instructions with corresponding inline assembly implementations.
This approach preserves the security mechanisms while ensuring successful compilation.

\textbf{Resolving mismatches in function prototypes.}
Decompilers struggle to accurately infer function prototypes due to compiler
optimizations and information loss, causing compilation failures when argument
counts and return types differ between call sites and callees. We observe that,
given our minimally invasive approach, if we consistently overestimate
arguments and return types (never marking non-void as void), the issue can be
resolved without affecting the overall functionality. However, underestimating them
can cause serious issues. For example, consider a case of a recompiled call site
calling an unmodified function. If we underestimate the arguments at the call site,
the callee will operate on arbitrary register values, causing unpredictable
behavior. Similarly, if an unaltered function calls our recompiled function
and expects a return value, but the decompiler erroneously marks the callee as
void, then the caller will operate on an uninitialized return value, leading to unpredictable behavior.
%

To implement an effective overestimation strategy (similar to TypeArmor~\cite{typearmor}),
we developed decompiler-specific approaches. For Hex-Rays, we decompile callees
before callers to produce overestimated argument counts, because it
conservatively interprets any argument register or stack value used before
initialization as a potential parameter when analyzing a function in isolation
without the context of call sites.
For Ghidra, decompiling functions in address order yields better results, as decompiling callees first causes external functions to be analyzed with incomplete signatures that propagate throughout the call tree.
Furthermore, Ghidra presents additional challenges, as it generates inconsistent function declarations---a function might be declared with a void return type but have its return value assigned to a non-void type at call sites, or the same function might have different signatures across multiple call sites.
Traditional solutions fail to resolve these inconsistencies: unifying all call sites to a single prototype requires determining which signature is correct without runtime information, while ignoring mismatches entirely would break compilation.
To address this, \system modifies Clang's AST traversal to dynamically fix mismatches during compilation. When detecting a type mismatch during cast analysis, we construct a new function type specific to that call expression, updating only the local call site's type representation while preserving the global function declaration. This per-call-site strategy enables each invocation to have its own effective signature without affecting other call sites.

For argument count mismatches where call sites pass more arguments than function declarations specify, we disable Clang's strict parameter count validation. Since arguments are passed via standard calling conventions and the callee accesses only those it requires, any excess arguments are safely ignored.
These targeted modifications
preserve the original binary's behavior while ensuring successful recompilation.

%

\subsubsection{Metadata Extractor} \label{metaextract}
The metadata extractor is responsible for extracting critical information from the original binary and producing metadata regarding the decompiled code.
These metadata include:
\begin{enumerate}
  \item Function boundaries and entry points, required by the patcher component for determining where to place trampolines.
  \item Layout of local variables on the stack, required by the recompiler component for
    fixing memory layout distortions. We extract the layout in the form of
    offsets from the frame pointer (e.g., \texttt{RBP}). For functions not using a frame pointer, we
    pick a pivot point (usually the first local variable/object on the stack). We then
    obtain offsets of all other objects with respect to this pivot point.
  \item Addresses of global data objects and external functions needed for symbol resolution and linking recompiled code with the
    original binary.
  \item Global typedefs, macros, function prototypes, and other definitions required by the recompiler component for fixing syntax issues and ensuring successful compilation.
\end{enumerate}
%

\subsection{Recompiler}
\label{sec:design_recompiler}
The recompiler component leverages the LLVM infrastructure through Clang plugins, LLVM
passes, and LLVM machine passes to implement binary-aware recompilation that 
applies semantic fixes and produces the compiled
object file. 
It consists of the following subcomponents:

\subsubsection{Stack Layout Preserver} \label{stack_layout_preserver}

This subcomponent primarily focuses on solving Challenge~\circled{2} (memory layout
distortion). It leverages customized LLVM passes and Clang plugins to
force the recompiled functions to have the stack layout obtained from the
decompiler (Section~\ref{metaextract}). This is done in two parts:

We first implemented a fix for boolean type sizes. The C standard does not
strictly define the size of boolean types, only specifying that they must be
large enough to store the values 0 and 1. This leads to compiler-dependent
implementations with varying boolean sizes. When decompilers identify local
variables as boolean types, recompilation issues can arise if the compiler uses
a larger size than in the original binary. Such size discrepancies can cause
stack objects to overlap due to different padding, leading to memory corruption.
To address this issue, we developed a Clang plugin that forces all boolean types
to be 1 byte in size, matching the most common implementation in original
binaries. 

Second, enforcing decompiler-extracted stack offsets requires addressing a fundamental challenge in LLVM's compilation pipeline. Stack frame layout occurs in the \texttt{PrologEpilogInserter} (PEI) pass, which runs after register allocation. At this stage, local variables exist only as frame indices—abstract references without concrete memory offsets. PEI's \texttt{calculateFrameObjectOffsets} function then assigns actual stack offsets based on variable sizes, alignment requirements, and calling conventions. Simply modifying offsets after this calculation is too late: subsequent passes have already generated instructions assuming the calculated layout.

Our solution exploits LLVM's fixed stack object mechanism, originally designed for stack objects with predetermined offsets, such as stack-passed function arguments and ABI-required slots. Unlike normal stack objects, fixed stack objects have predetermined offsets that LLVM's backend must preserve. We intercept PEI before \texttt{calculateFrameObjectOffsets} executes, converting decompiler-identified variables from normal stack objects (frame indices only) to fixed stack objects with their original binary offsets. This transformation occurs through custom machine function passes that: (1) parse decompiler metadata to map variable names to stack offsets, (2) match LLVM's debug information against this mapping to identify corresponding frame indices, and (3) invoke \texttt{MachineFrameInfo::CreateFixedObject} to instantiate fixed stack objects at the specified offsets. Because fixed stack objects integrate directly into LLVM's frame lowering infrastructure, this approach works seamlessly across all supported architectures without backend-specific modifications.

However, this creates a constraint satisfaction problem: we must honor the original binary's exact offsets while satisfying LLVM's alignment requirements for correct code generation. For example, SSE instructions like \texttt{movaps} mandate 16-byte alignment; accessing unaligned addresses can trigger segmentation faults. When the original binary places a variable at offset -0x14 (not 16-byte aligned) and the recompiled code uses SSE operations, we cannot simply preserve the original offset. Our implementation resolves this through hybrid allocation: variables identified in decompiler metadata retain their exact original offsets, while additional variables introduced during recompilation receive calculated offsets that satisfy alignment constraints through LLVM's standard frame layout algorithm. Algorithm~\ref{alg:stack_layout} summarizes this process.

\begin{algorithm}[!ht]
\caption{Binary-Aware Stack Layout}\label{alg:stack_layout}
\small
\begin{algorithmic}[1]
\STATE \textbf{Input:} MachineFunction $F$, metadata $M$ (var $\to$ offset)
\STATE $\textit{layout} \gets$ Parse stack layout from $M$ for function $F$
\STATE \textcolor{green!60!black}{// $\textit{layout}$ maps variable names $v$ to stack offsets $\textit{target}$}
\STATE
\STATE \textcolor{green!60!black}{// For Stack Pointer (SP) based functions,}
\STATE \textcolor{green!60!black}{// convert offsets to Base Pointer (BP) based convention}
\IF{$F$ originally SP-based}
    \STATE $\textit{max} \gets \max_{v}(\text{size}(v) + \textit{layout}[v])$
    \STATE $\textit{layout}[v] \gets \textit{layout}[v] - \textit{max}$ for all $v$
\ENDIF
\STATE
\STATE \textcolor{green!60!black}{// Prevents conflicts with original variable offsets}
\STATE Create fixed stack objects for callee-saved register spills
\STATE
\STATE \textcolor{green!60!black}{// Fix original variables at exact binary offsets}
\FORALL{$(v, \textit{target}) \in \textit{layout}$}
    \IF{$\exists$ frame index $i$ with name $v$}
        \IF{no fixed allocation at $\textit{target}$}
            \STATE Allocate fixed slot at $\textit{target}$ with alignment = 1
            \STATE Replace all uses of $i$ with fixed slot
        \ENDIF
    \ENDIF
\ENDFOR
\STATE
\STATE \textcolor{green!60!black}{// Compiler-introduced variables; uses built-in allocation with proper alignment}
\STATE Allocate new variables via LLVM's standard frame layout
\end{algorithmic}
\end{algorithm}

Our implementation addresses an LLVM-specific detail: the backend's frame reference calculation applies an internal transformation to stack offsets. We compensate by adjusting the declared offset such that the final calculated offset matches our target. This approach preserves the exact memory layout of stack objects, ensuring compatibility with unaltered components of the original binary. Beyond struct recovery, this mechanism correctly handles stack canaries, register spill slots (created during register allocation, assigned offsets during PEI), and calling convention mismatches (where decompilers misidentify argument passing mechanisms). Each of these cases depends on precise stack layout preservation—a single byte of offset error causes memory corruption or control-flow hijacking.

\textbf{Concrete Example: Struct Decomposition.}
Recall the struct decomposition issue from Challenge~\circled{2} (Section~\ref{memlayoutissue}), illustrated in Listing~\ref{lst:b2sum_blake2b_init} where Hex-Rays decompiles a 64-byte \texttt{blake2b\_param} struct as seven separate char arrays (\texttt{v4} through \texttt{v10}).

With \system's stack layout preservation, the recompiled code is identical to the decompiled code in Listing~\ref{lst:b2sum_blake2b_init}, with each variable placed at the same offset, preserving the contiguous 64-byte memory layout. When this data is passed to \texttt{blake2b\_init\_param} as a struct pointer, all bytes are correctly initialized.

Without \system (Listing~\ref{lst:b2sum_nofixed}), the compiler reorders variables arbitrarily (\texttt{v17} at \texttt{[rbp-0x4]}, \texttt{v7} at \texttt{[rbp-0x3C]}), breaking the contiguous layout. Only the first four bytes are correctly initialized, causing subsequent hash computation failures.

\begin{listing}[!ht]
\begin{minted}{c}
int64 blake2b_init(int64 a1, unsigned int64 a2)
{
  char v3[24]; // [rsp+0h] [rbp-70h] BYREF
  unsigned int64 v4; // [rsp+18h] [rbp-58h]
  char v5[14]; // [rsp+20h] [rbp-50h] BYREF
  char s[6]; // [rsp+2Eh] [rbp-42h] BYREF
  char v7[4]; // [rsp+34h] [rbp-3Ch] BYREF
  char v8[4]; // [rsp+38h] [rbp-38h] BYREF
  char v9[4]; // [rsp+3Ch] [rbp-34h] BYREF
  unsigned int64 v10; // [rsp+40h] [rbp-30h]
  int64 v11; // [rsp+48h] [rbp-28h]
  int v13; // [rsp+5Ch] [rbp-14h]
  unsigned int64 v14; // [rsp+60h] [rbp-10h]
  char v15; // [rsp+6Ah] [rbp-6h]
  char v16; // [rsp+6Bh] [rbp-5h]
  char v17[4]; // [rsp+6Ch] [rbp-4h] BYREF

  v11 = a1;
  v14 = a2;
  v13 = 40;
  v10 = __readfsqword(0x28u);
  v4 = v10;
  if ( !a2 || v14 > 0x40 )
    return 0xFFFFFFFFLL;
  v17[0] = v14;
  v17[1] = 0;
  v17[2] = 1;
  v17[3] = 1;
  store32(v7, 0LL);
  store32(v8, 0LL);
  store32(v9, 0LL);
  v15 = 0;
  v16 = 0;
  memset(s, 0, 0xEuLL);
  memset(v5, 0, 0x10uLL);
  memset(v3, 0, 0x10uLL);
  return blake2b_init_param(v11, v17);
}
\end{minted}
\caption{Recompiled without \system, breaking memory layout}
\label{lst:b2sum_nofixed}
\end{listing}

\subsubsection{Direct Symbol Enforcer}
To prepare for symbol resolution in the patcher component, we ensure recompiled objects are suitable for direct linking (to address Challenge~\circled{3}). For references to other functions or global variables in the original binary, we mark them as \texttt{dso\_local}~\cite{llvmlangref} using an extra LLVM pass. This tells the compiler that these symbols will be resolved within the same linkage unit, forcing it to generate code that directly accesses these symbols, rather than relying on external linking mechanisms (e.g., GOT entries). As we will explain, the patcher component will later fix the actual addresses of these symbols.

\subsection{Patcher}

The patcher component retrofits the program binary with newly recompiled
functions, addressing Challenge~\circled{3}. It acts as a specialized linker
that takes the object file containing recompiled functions as input and links it
with the unmodified components of the original binary. We have built our patcher
component on top of Patcherex2~\cite{patcherex2}, a generic binary patching tool
that supports multiple architectures, file formats, and compiler toolchains. It
has three main subcomponents:

\subsubsection{Symbol Resolver}
As \system only recompiles specific functions, it needs to ensure that the
recompiled functions can correctly refer to functions and global variables in the
original binary (Challenge~\circled{3}). This is crucial for maintaining
the original binary's behavior, ensuring that the recompiled functions
can interact seamlessly with the rest of the binary.

\textbf{Linking Without Object Files.}
Standard linking requires object files for all code being linked. However,
\system only recompiles specific functions; the rest of the binary remains in
compiled form with no corresponding object files. This prevents traditional
linking: the linker cannot resolve references to functions and data in the
unmodified binary portions.

We address this by dynamically generating custom linker scripts. The script explicitly maps each symbol name to its address in the original binary (e.g., \texttt{printf = 0x401050;}), using the symbol-to-address mapping
extracted via decompiler metadata (Point 3 in Section~\ref{metaextract}).
This allows the linker to resolve references in recompiled code directly to their binary addresses
without requiring object files for unmodified code. References to external
(third-party library) functions and objects are directed to the original GOT
table. 
\subsubsection{Binary Retrofitter}
After resolving symbols, the patcher component retrofits the recompiled function
into the original binary. The recompiled function may be larger due to the
user-applied patch. To accommodate it, we cannot relocate other functions or
data, as doing so would disturb the control flow and data access in the unaltered
components of the binary. Therefore, we follow a layout-preserving approach.
First, our patcher component attempts in-place replacement when the recompiled
function is smaller. If that is not possible, it tries to utilize existing
binary padding, and finally, if necessary, it modifies the binary header to
create a new segment.

\begin{figure}
  \centering
  \includegraphics[width=1\linewidth, trim=3pt 5pt 0pt 0pt, clip]{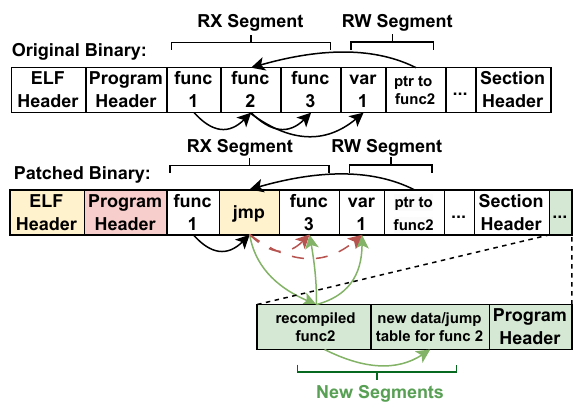}
  \caption{Overview of our binary rewriting technique for preserving symbolic
  references}
  \label{fig:elf_rewrite}
\end{figure}

Figure~\ref{fig:elf_rewrite} illustrates our layout-preserving binary rewriting
approach, where red elements indicate deletions, green indicates additions, and
yellow represents modifications. When patching a function (e.g.,
\texttt{func2}), \system creates space for the recompiled version without
modifying the original code and data layout. Note that the function might be
accessed by other functions using a function pointer (e.g., \texttt{ptr} to
\texttt{func2} in Figure~\ref{fig:elf_rewrite}). Since identifying pointers is
hard (Challenge~\circled{3}), \system does not attempt to modify them. Instead,
it inserts a trampoline {\em jmp} instruction at the original function's
location, directing execution to the recompiled version. Direct calls from
other functions are also redirected via this trampoline. 


This approach ensures the correctness and transparency of the
binary-aware recompilation. The framework itself introduces minimal overhead:
only a trampoline jump when recompiled functions are enlarged beyond their
original size. Since we recompile targeted functions rather than whole binaries,
execution overhead depends on the specific patch and approximates source-level
instrumentation. This contrasts with other approaches~\cite{prd} that use jump
tables or wrapper functions for every instrumented function.

%
%
%
%
%

\subsubsection{Additional Data Handler}

The recompiled binary may contain additional data beyond just function
code. Common examples include string literals, jump tables for
switch-case statements, and additional global variables. These elements are
crucial for the recompiled function.

For jump tables in switch-case statements, we cannot use the old jump tables in
the original binary because modifying and aligning them with newly
recompiled case blocks can be hard. Therefore, we allow the compiler to generate new jump
tables based on the decompiled switch-case statements. Our binary retrofitter
injects this new jump table into the binary and places it at the right offset
from the recompiled function. This placement preserves low-level
arithmetic involved in accessing jump table entries and thereby preserves the
function's behavior.  

For string literals, global variables, and other non-code elements, we follow
a process similar to function retrofitting, finding or creating appropriate
segments and updating the binary header accordingly.
Handling both code and non-code
elements through our binary-aware recompilation ensures that all
necessary data is either preserved or accurately reconstructed in the recompiled
version, maintaining semantic equivalence.

\subsection{Implementation}

\system integrates multiple components through a Python-based orchestration layer.
The decompiler integration supports both Hex-Rays and Ghidra,
extracting function code, metadata, and symbol information.
The compiler modifications include a patch to Clang 15's semantic analysis at the AST level for handling function prototype mismatches, custom IR passes applied after C-to-IR lowering, and modifications to the PrologEpilogInserter pass in the machine code generation stage after register allocation for stack layout preservation using binary metadata.
All components work together through a unified pipeline that processes decompiled code,
applies fixes, and generates binary-compatible object files.

Rather than detecting specific issues in each function, \system applies all fixes to each recompiled function.
These fixes are designed to be conservative transformations that preserve correctness even when not strictly necessary. For example, stack layout preservation uses binary metadata to enforce exact offsets; if the compiler would naturally choose the same layout, the fix has no effect. Similarly, syntax fixes like normalizing malformed names or adding type definitions do not alter semantics when the original code was already correct. This universal application strategy eliminates the need for issue detection heuristics and ensures comprehensive coverage.

For large-scale evaluation, we deployed \system on a distributed Kubernetes infrastructure with message queues for task distribution, processing tens of thousands of function-level recompilation tasks across multiple optimization levels.
This architecture enables flexible scaling to handle comprehensive evaluation workloads across entire datasets.


\section{Evaluation}
\label{sec:eval}
In this section, we first assess \system's ability to recompile functions and correct semantic inaccuracies.
Next, we evaluate its real-world applicability by using it to patch real-world vulnerabilities. 
Finally, we conduct a study involving humans, as well as LLMs, to determine whether it simplifies the process of applying patches to decompiled code.
Overall, we aim to answer the following four key research questions:

\begin{itemize}
    \item RQ1: (Recompilability) How often can the decompiled code, produced by existing decompilers, be recompiled using only \system's syntactic fixes? -- (Section~\ref{recompeval})
    \item RQ2: (Semantic Correctness) How effective is \system at addressing semantic inaccuracies in the recompiled code? -- (Section~\ref{semeval})
    \item RQ3: (Real-world Applicability) Can \system be used to fix real-world vulnerabilities by enabling patching of decompiled code? -- (Section~\ref{realworldeval})
    \item RQ4: (Usability) How easy is it for both humans and LLMs to use \system to patch decompiled code? -- (Section~\ref{usability})
\end{itemize}

\subsection{Experimental Setup}
For RQ1 and RQ2, we evaluated \system's effectiveness on two datasets: (i)
Coreutils~\cite{coreutils} (v9.4) and (ii) DARPA CGC
challenges~\cite{cgc_challenges} binaries. Coreutils was compiled using GCC
11.4.0 on Ubuntu 22.04, while CGC binaries used GCC 8.4.0 on
Ubuntu 20.04. Both datasets were compiled with \texttt{-O0} through \texttt{-O3}
optimization flags. For decompilation, we used IDA Pro 8.4 with the Hex-Rays
decompiler plugin and Ghidra 11.2.
We selected the CGC dataset for its diverse application types, representing various software categories with different behaviors and structures. Coreutils was chosen because it represents fundamental Linux tools, providing a comprehensive and widely deployed codebase for evaluation. Critically, both datasets include comprehensive test suites with high code coverage, enabling reliable validation of recompiled binaries. This combination tests \system across specialized applications and essential system utilities.

To evaluate \system at scale, we designed our test infrastructure
to run on a Kubernetes cluster with 2,000 cores and 8,000 GiB of physical memory.
The system comprises three main components: a message queue dispatching workloads to worker pods; worker pods (2 cores, 8 GiB memory each) running \system components and executing test cases; and an object storage system storing all intermediate and final results.

We conducted a large-scale evaluation of \system by testing each function of each binary in our dataset individually. Specifically, each function was decompiled, then recompiled and patched back into the binary, resulting in as many versions of
a given binary as the total number of functions. We then ran all dataset test cases on each recompiled binary to evaluate correctness, enabling fine-grained evaluation of the recompilation correctness for each individual function.

Our evaluation follows a multi-stage pipeline. Starting from the total functions in each dataset (TF), we first apply \system's syntax fixes (Section~\ref{syntax_fixer}). Functions that compile successfully (Recomp) enter the testing phase. We categorize compiled functions into two groups: those that pass all test cases without semantic fixes (TestPass), indicating correct decompilation, and those that fail tests (SemError), revealing semantic errors such as incorrect memory layouts or type mismatches. For functions with semantic errors, we apply \system's semantic fixes (Section~\ref{sec:design}) and re-test to determine how many are successfully fixed (FixedSem). Table~\ref{tab:recompilation_results} presents results at each stage, with all percentages calculated relative to the total function count unless otherwise specified.

\begin{table*}
    \centering
    \caption{Recompilation results with Ghidra and Hex-Rays. \system achieves
    successful recompilation for 84.1\% of Hex-Rays and 49.4\% of Ghidra
    functions (Recomp). Semantic errors occur in 17.9\% (Hex-Rays) and 6.1\%
    (Ghidra) of functions (SemError), of which \system fixes 80.9\% and 53.3\%
    respectively (FixedSem over SemError).}
    \label{tab:recompilation_results}
    \begin{adjustbox}{width=\textwidth}
    \tabcolsep 2pt
    \begin{tabular}{|c|c|c|@{\vrule width 1pt}|c|c|c|c|c|@{\vrule width 1pt}|c|}
        \hline
        \multirow{2}{*}{Decompiler} & \multirow{2}{*}{Dataset} & \multirow{2}{*}{Opt Lvl.} & Total Functions & Recompiled Funcs. & Passing Tests w/o Fixes & Failing Tests w/o Fixes & \system Fixed Funcs. & \system Fixed \% \\
        ~ & ~ & ~ & (TF) & (Recomp) & (TestPass) & (SemError) & (FixedSem) & (FixedSem over SemError) \\
        \hline
        \hline
        \multirow{8}{*}{Hex-Rays} & \multirow{4}{*}{Coreutils} & -O0 & \num{18775} & \num{16977} (90.4\%) & \num{14532} (77.4\%) & \num{2445} (13.0\%) & \num{2209} (11.8\%) & 90.3\% \\
        \cline{3-9}
        ~ & ~ & -O1 & \num{17249} & \num{14933} (86.6\%) & \num{12553} (72.8\%) & \num{2380} (13.8\%) & \num{1934} (11.2\%) & 81.3\% \\
        \cline{3-9}
        ~ & ~ & -O2 & \num{14504} & \num{10618} (73.2\%) & \num{8294} (57.2\%) & \num{2324} (16.0\%) & \num{1750} (12.1\%) & 75.3\% \\
        \cline{3-9}
        ~ & ~ & -O3 & \num{14575} & \num{10026} (68.8\%) & \num{7868} (54.0\%) & \num{2158} (14.8\%) & \num{1635} (11.2\%) & 75.8\% \\
        \cline{2-9}
        ~ & \multirow{4}{*}{CGC} & -O0 & \num{4416} & \num{4119} (93.3\%) & \num{3256} (73.7\%) & \num{863} (19.5\%) & \num{755} (17.1\%) & 87.5\% \\
        \cline{3-9}
        ~ & ~ & -O1 & \num{3882} & \num{3581} (92.2\%) & \num{2724} (70.2\%) & \num{857} (22.1\%) & \num{755} (19.4\%) & 88.1\% \\
        \cline{3-9}
        ~ & ~ & -O2 & \num{3730} & \num{3277} (87.9\%) & \num{2401} (64.4\%) & \num{876} (23.5\%) & \num{666} (17.9\%) & 76.0\% \\
        \cline{3-9}
        ~ & ~ & -O3 & \num{3801} & \num{3059} (80.5\%) & \num{2284} (60.1\%) & \num{775} (20.4\%) & \num{566} (14.9\%) & 73.0\% \\
        \hline
        \multicolumn{3}{|c|@{\vrule width 1pt}|}{Hex-Rays Average} & N/A & 84.1\% & 66.2\% & 17.9\% & 14.5\% & 80.9\% \\
        \hline
        \hline
        \multirow{8}{*}{Ghidra} & \multirow{4}{*}{Coreutils} & -O0 & \num{19314} & \num{13086} (67.8\%) & \num{11466} (59.4\%) & \num{1620} (8.4\%) & \num{777} (4.0\%) & 48.0\% \\
        \cline{3-9}
        ~ & ~ & -O1 & \num{20289} & \num{10816} (53.3\%) & \num{9911} (48.8\%) & \num{905} (4.5\%) & \num{548} (2.7\%) & 60.6\% \\
        \cline{3-9}
        ~ & ~ & -O2 & \num{19735} & \num{8984} (45.5\%) & \num{7984} (40.5\%) & \num{1000} (5.1\%) & \num{526} (2.7\%) & 52.6\% \\
        \cline{3-9}
        ~ & ~ & -O3 & \num{19909} & \num{8947} (44.9\%) & \num{7887} (39.6\%) & \num{1060} (5.3\%) & \num{540} (2.7\%) & 50.9\% \\
        \cline{2-9}
        ~ & \multirow{4}{*}{CGC} & -O0 & \num{4686} & \num{2747} (58.6\%) & \num{2388} (51.0\%) & \num{359} (7.7\%) & \num{192} (4.1\%) & 53.5\% \\
        \cline{3-9}
        ~ & ~ & -O1 & \num{4613} & \num{2213} (48.0\%) & \num{1893} (41.0\%) & \num{320} (6.9\%) & \num{168} (3.6\%) & 52.5\% \\
        \cline{3-9}
        ~ & ~ & -O2 & \num{4693} & \num{1846} (39.3\%) & \num{1590} (33.9\%) & \num{256} (5.5\%) & \num{138} (2.9\%) & 53.9\% \\
        \cline{3-9}
        ~ & ~ & -O3 & \num{4760} & \num{1793} (37.7\%) & \num{1539} (32.3\%) & \num{254} (5.3\%) & \num{138} (2.9\%) & 54.3\% \\
        \hline
        \multicolumn{3}{|c|@{\vrule width 1pt}|}{Ghidra Average} & N/A & 49.4\% & 43.3\% & 6.1\% & 3.2\% & 53.3\% \\
        \hline
    \end{tabular}
    \end{adjustbox}
\end{table*}

\subsection{RQ1 -- Recompilability of decompiled code} \label{recompeval}
Our analysis of the results presented in Table~\ref{tab:recompilation_results} reveals that, after applying \system's syntax fixes (Section~\ref{syntax_fixer}), decompiled code produced by Hex-Rays achieves 84.1\% recompilability success. This demonstrates \system's effectiveness in overcoming fundamental syntactic barriers to binary recompilation. In contrast, decompiled code produced by Ghidra achieves a 49.4\% recompilation rate after syntax fixes. This is largely due to Ghidra's less strict adherence to C language standards, resulting in a wide variety of syntactic issues such as non-standard type representations (e.g., 7-bit \texttt{int7} types). While each individual issue may be addressable, the large diversity of such deviations collectively demands substantial engineering effort. Our current implementation addresses the most common cases, but exhaustive handling of all edge cases remains impractical. Although Ghidra often preserves semantic meaning, addressing these varied syntactic deviations necessitates incremental decompiler-specific enhancements as new patterns emerge.

\subsection{RQ2 -- Fixing semantic incorrectness using \system} \label{semeval}
\system effectively resolves a substantial portion of semantic issues in recompiled functions that initially failed test cases. Across all optimization levels, it fixes 80.9\% of semantic errors in Hex-Rays and 53.3\% in Ghidra decompiled code (Table~\ref{tab:recompilation_results}). Notably, at \texttt{-O2}—the standard optimization level for most real-world programs—\system corrects 75.7\% of semantic issues in Hex-Rays outputs.

As described in Section~\ref{sec:design}, \system applies all fixes universally to every function. The 80.9\% semantic fix rate represents functions where at least one semantic fix was necessary to achieve correct behavior, though many functions benefit from multiple fixes simultaneously as noted in Section~\ref{sec:challenges}.

\textbf{Interdependence of Fixes.}
Traditional ablation studies that disable individual components to measure their isolated contributions are not applicable to \system's fix pipeline. As established in Section~\ref{sec:challenges}, functions typically exhibit multiple overlapping issues: a single function may require multiple fixes simultaneously. Disabling any single fix would cause the function to fail due to remaining unaddressed issues, preventing isolation of individual contributions. Our evaluation instead demonstrates that the complete pipeline is necessary: without \system's fixes, only 66.2\% of Hex-Rays functions pass tests, whereas with all fixes applied, this increases to 80.7\% (66.2\% baseline + 14.5\% semantic fixes).

\begin{figure}
    \includegraphics[width=1\linewidth, trim=10pt 10pt 0pt 0pt, clip]{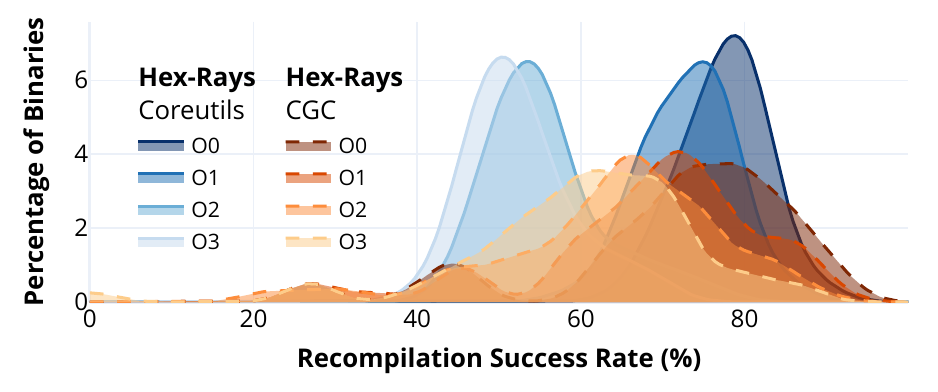}
    \includegraphics[width=1\linewidth, trim=10pt 10pt 0pt 0pt, clip]{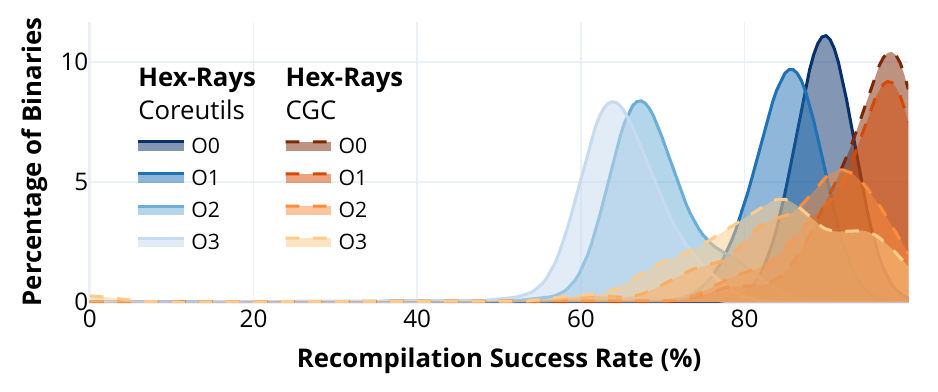}
    \caption{Hex-Rays Recompilation Success Rate before (top) and after (bottom) applying \system's fixes}
    \label{fig:recomp_success_rate}
\end{figure}

To complement our function-level results, we also examined recompilation
coverage at the binary level. Figure~\ref{fig:recomp_success_rate} shows the
distribution of recompilation outcomes using the Hex-Rays decompiler across all
optimization levels for both datasets. 

Before applying \system (top panel), many binaries fail to recompile more than half of their functions.
In contrast, after applying \system (bottom panel),
the success rate distributions shift markedly to the right.
Specifically, across all optimization levels 
and datasets, the number of cases achieving over 80\% recompilation success increases from 
80 to 394 (out of 640), while cases below 60\% drop from 235 to just 7. Notably, after applying \system, 
we achieve complete recompilation (100\% success) in the CGC dataset for 12 (out of 54) binaries at \texttt{O0}, 
8 binaries at \texttt{O1}, and 2 binaries each at \texttt{O2} and \texttt{O3}. Beyond these complete successes, 
56 additional CGC binaries achieve greater than 90\% recompilation success. While \system is designed primarily 
for targeted patching, these successful cases of full or near-full binary recompilation suggest potential 
for future applications such as whole-binary hardening and instrumentation.


\subsubsection{Analysis of Remaining Failures}

While \system fixes 80.9\% of semantic errors in Hex-Rays output, the remaining 19.1\% expose deeper decompiler challenges. This section highlights the most common error patterns. Although not exhaustive, these patterns capture analysis gaps that lie beyond recompilation-time fixes and are best addressed through improvements to the decompiler itself.


\textbf{Multi-Register Value Handling.} Decompilers sometimes misidentify return values spanning multiple registers. For example, on x86-64 systems, structs containing two 8-byte types should return values across \texttt{rax} and \texttt{rdx} registers according to the calling convention. However, Hex-Rays may incorrectly assume only \texttt{rax} holds the return value, treating \texttt{rdx} as a scratch register and losing half the return value in the decompiled code.

\textbf{Type Inference Inaccuracies.} Some functions exhibit significant type inference errors where decompilers produce types that are incorrect even at the fundamental level (e.g., wrong sizes). These pervasive type errors affect numerous variables and operations within a function, making targeted fixes impractical without addressing the root cause in the decompiler's type analysis.

\textbf{Optimization-Level Degradation.} Fix rates decline at higher optimization levels (-O2: 75.3\%, -O3: 75.8\% for Hex-Rays) due to aggressive compiler transformations that make decompiler type recovery less accurate. However, even at -O3, \system maintains over 75\% fix rate, demonstrating robustness across optimization levels.



\subsection{RQ3 -- Real-world Applicability} \label{realworldeval}
To evaluate the applicability of \system, we manually examined all CVEs reported in GNU Coreutils~\cite{coreutils} over the past 10 years and GNU Binutils~\cite{binutils} over the past 5 years. From this set, we identified 14 CVEs (2 in Coreutils, 12 in Binutils) that have publicly available, easily reproducible proof-of-concept (PoC) exploits that cause crashes with user-controlled input. We were able to successfully reproduce all 14 of these vulnerabilities using the published PoC exploits. Other reported vulnerabilities either lacked public PoCs, were not directly exploitable, or required specific environments to trigger, making them less suitable for our evaluation.

We applied \system to all 14 reproducible CVEs and successfully patched 13 of
them. The patched binaries passed all test cases while rendering the PoC exploit
ineffective. The remaining CVE was not patched successfully due to a function-pointer
corner case for which we have not yet implemented a fix, given the rarity of the issue.

In total, we recompiled 15 functions across the 14 CVEs, with one CVE requiring
patches in two distinct functions. The size and complexity of the patched
functions varied significantly, ranging from less than 100 lines of code to over
2,000 lines (average of 379 lines and a median of 277 lines). This distribution
demonstrates \system's ability to handle functions of various sizes and
complexities, from relatively simple routines to large, intricate functions
with complex control flow and data structures. 
For all 14 CVEs, the process of developing and applying the necessary fix in the
decompiled code took less than 5 minutes per vulnerability, given access to the
original diff file. Notably, we did not need to manually modify anything in the
decompiled code beyond applying the specific fix for each vulnerability. This
shows that minimal manual effort is required to apply patches using \system,
highlighting the practicality of the tool. 
We further evaluated automated patch development by using LLMs
to generate patches for decompiled CVE fixes, using source diff patches as references. While some
generated patches required minor manual adjustments,
they were functionally equivalent to our manually created patches,
highlighting the feasibility of automated patch development with \system.

By successfully patching real-world vulnerabilities, \system proves to be a robust tool
for fixing security issues in binary software, even when dealing with large and
complex codebases.
Appendix~\ref{appendix:cve_details} lists all evaluated CVEs with their characteristics.
In Section~\ref{sec:case}, we present detailed case studies of selected security
vulnerabilities successfully patched by \system.

\subsection{RQ4 -- Usability}
\label{usability}
To objectively evaluate the practical usability of \system for applying patches to decompiled code, we conducted a controlled user study with 18 participants and automated evaluations with three state-of-the-art Large Language Models (LLMs). Each participant completed a 3.5-hour session consisting of 6 tasks based on 3 vulnerabilities selected from the 13 successfully patched CVEs described in Section~\ref{realworldeval}.

\textbf{Participant Recruitment.}
We conducted a comparative user study to evaluate the practicality of \system.
To this end, we obtained approval from our university's Institutional Review Board (IRB) and recruited participants through advertisements in internal department groups, classroom announcements, and flyers on bulletin boards.
To ensure we had participants with adequate experience with decompilers and memory-related vulnerabilities, we developed a screening questionnaire with questions related to their experience with decompilers, debugging, and C code.
We excluded participants from the user study if they indicated they had no understanding of basic C concepts like pointers and type casts or were unable to read git diff results. However, we did not require prior experience with decompilers or memory-related vulnerabilities.
In total, we recruited 18 participants.

\textbf{Study Design.}
We selected three vulnerabilities of varying difficulty levels (easy, medium, and hard) from our successfully patched CVEs. The difficulty classification was based on two key factors: the complexity of locating where to insert the patch and the complexity of formulating the correct patch. This classification reflects the challenges a human operator may face in navigating and modifying decompiled code, rather than the difficulty \system may encounter in addressing the underlying semantic issues.
The easy vulnerability was a simple typo requiring a global variable name change
to ensure proper cleanup. While straightforward to patch, recompilation of the
decompiled code would cause stack corruption without our stack layout
preservation technique (Section~\ref{stack_layout_preserver}). The medium
vulnerability required removing redundant buffer reallocation code, presenting
moderate difficulty in identifying and removing unnecessary memory operations. The
hard vulnerability required adding a null check before dereferencing a pointer,
which is challenging because it necessitates identifying pointer arithmetic equivalent to
source-level member access. This also triggers stack corruption without
\system's semantic fixes.
Each participant was asked to patch each CVE twice: once with \system and once without it. To mitigate learning effects from repeated patching attempts, we randomized both the order of CVEs and whether \system was used first or second for each vulnerability. Each task was limited to 30 minutes.

Participants worked in a controlled environment with a browser-based Ubuntu 22.04 VM that had development and binary analysis tools pre-installed (VS Code, Ghidra, GDB, vim) and Internet access.
For each task, participants were provided with the original binary, source code patch in the form of a \texttt{diff} file, and the vulnerable function's decompiled code. They were also given a patching script that recompiles their patched code and retrofits it into the original binary, and a test script to validate patched binaries against the original test case.

The study was conducted over Zoom, allowing us to observe participants' behavior and problem-solving approaches throughout the session.
During the study, the participants were given full freedom to utilize any resources they needed to achieve their task, including the ability to install other tools and access online resources.
At the end of the study, the participants were asked to fill out a survey
regarding their experience with patching the vulnerabilities. This survey asked
participants to rate the difficulty level of each task, the challenges they
faced for each task, and what additional information they required to apply the
patches.
To identify potential issues in our study design, we preliminarily conducted a pilot study with two developers.
Based on their feedback, we refined our study by clarifying task instructions, adding questions to the questionnaires, and improving the usability of our evaluation setup.

In parallel, we evaluated three LLMs (GPT-5, Claude 4.5 Sonnet, Gemini 2.5 Pro) on the same three vulnerabilities using three trials per vulnerability per condition (with/without \system). LLMs received identical materials as human participants: decompiled code, source patch diff, and test scripts. To enable fair comparison, we configured each LLM as an agent with tool-calling capabilities, exposing five functions: \texttt{propose\_patch} (to submit patch proposals), \texttt{decompile\_function} (to invoke Ghidra headless decompilation), \texttt{run\_command} (to execute arbitrary shell commands including GDB scripts, objdump, readelf, and custom analysis scripts), \texttt{run\_test} (to validate patches), and \texttt{done} (to complete the task). This function calling interface allowed LLMs to programmatically invoke debugging tools and iteratively refine patches based on test failures, mirroring the debugging strategies available to human participants. We allowed up to 30 tool-based patch attempts per task with a 30-minute timeout. This setup tests whether LLMs can overcome semantic barriers through extensive debugging and iteration, similar to how human experts used GDB and Ghidra.

\textbf{User Study Results.}
The results demonstrate a dramatic difference in participants' ability to successfully patch vulnerabilities with and without \system as shown in Table~\ref{tab:patching_results}.
By using \system, participants achieved a 100\% success rate across all
vulnerability difficulty levels. In contrast, without \system, participants
experienced minimal success when attempting to patch decompiled code directly.
Only 2 participants managed to successfully address the medium vulnerability.
The fact that no participants could patch even the easy vulnerability without
\system highlights that regardless of patch simplicity, underlying syntactic and
semantic issues in decompiled code prevent even experienced participants from
successfully implementing patches without \system.

\begin{table}[h]
    \centering
    \caption{Comparison of participants' success in patching vulnerabilities with and without \system}
    \label{tab:patching_results}
    \begin{tabular}{|l|c|c|}
        \hline
        \textbf{Vulnerability} & \textbf{W/O \system} & \textbf{With \system} \\
        \hline
        Easy Vulnerability    & 0/18  & 18/18 \\
        Medium Vulnerability  & 2/18  & 18/18 \\
        Hard Vulnerability    & 0/18  & 18/18 \\
        \hline
    \end{tabular}
\end{table}

For participants who were able to successfully patch the vulnerability, we also measured the time it took them to apply the patch.
Successful patch times also showed significant differences. With \system, participants typically completed patches in less than 5 minutes for the easy and medium vulnerabilities, and less than 10 minutes for the hard vulnerability. Without \system, most participants used the entire 30-minute allocation without success.
Even in the task where participants attempted to patch the hard vulnerability with \system first (eliminating any learning effect), it took participants an average of only 8 minutes and 29 seconds to successfully apply the patch. Detailed per-participant timing data for all three vulnerabilities is available in Appendix~\ref{appendix:study_time}.

With \system, participants completed patches in an average of 7 minutes. Without \system, only 2 participants succeeded; for all failed attempts, we conservatively assigned the maximum time (30 minutes) as the patch time. We performed a one-sided Wilcoxon signed-rank test~\cite{Woolson2008} with the null hypothesis that patch time with \system is greater than or equal to patch time without \system, and the alternative hypothesis that \system reduces patch time. The test yielded $p < 0.00001$, allowing us to reject the null hypothesis and conclude that \system significantly reduces time to successful patching.

In the post-study survey, all participants rated the tasks without \system as more difficult than those performed with it.
The most commonly cited challenge in the patching process was determining why their patch was incorrect.
Additionally, when asked about the difficulties they encountered, most participants reported no issues using \system.
However, without it, they struggled to identify the reasons their patches failed.
To apply patches without \system, participants had to manually \emph{``identify global variables that were causing problems''} and \emph{``identify location where the stack corruption happens''}.

We observed several key patterns in developer behavior when patching decompiled code.
Without \system, participants often experienced self-doubt, second-guessing initially correct patches after encountering compilation failures.
They also struggled with root cause identification, failing to recognize critical issues such as stack layout problems and missing decompiler-specific constructs.
Some explored alternative approaches, such as searching for original CVEs online, but faced challenges due to missing structure definitions.
Even participants who claimed to have advanced expertise in C programming failed to patch vulnerabilities without \system, emphasizing the inherent difficulty of the task.
Additionally, debugging strategies differed significantly.
Some participants relied heavily on tools like GDB and Ghidra but found them largely ineffective.
In contrast, when using \system, they rarely used such tools.
These observations highlight the significant barriers posed by semantic inaccuracies and demonstrate that \system substantially improves patching capabilities.

\textbf{LLM Results.}
Table~\ref{tab:llm_results} in Appendix~\ref{appendix:llm_results} presents detailed results for all three LLMs.
All three LLMs achieved 100\% success with \system (27/27 trials), while none succeeded without it (0/27). With \system, LLMs completed patches rapidly (12-40 seconds, succeeding in 1-2 attempts). Without \system, all models failed despite extensive debugging, exhausting 6-30 patch attempts before timeout.
These results demonstrate that \system enables fully automated binary patching workflows, opening new possibilities for large-scale vulnerability remediation in legacy software without human intervention.

LLM behavior closely mirrored human failure patterns: without \system, both groups (LLMs and human participants) initially developed semantically correct patches matching the source code fix. However, after test failures caused by decompilation artifacts, both abandoned correct approaches and attempted progressively worse variations. Critically, neither group could identify or address the root causes.
Overall, the findings from our user study indicate that \system enables the successful application of source-level patches, even for developers who are not experts in binary analysis, while also enabling automated patching approaches through LLM integration.


\section{Case Studies}
\label{sec:case}

We now present two cases from the 13 CVEs we fixed using \system (discussed in Section~\ref{realworldeval}), to demonstrate how \system can be used to patch real-world security vulnerabilities.

\textbf{CVE-2014-9471.} \label{cve-2014-9471}
This vulnerability~\cite{CVE-2014-9471} is a double-free issue in the GNU Coreutils \texttt{tail} program. While the official patch is straightforward at the source level, implementing it at the assembly level is challenging, as it modifies control flow by breaking a loop earlier. Listing~\ref{lst:cve-2014-9471} shows both the source and decompiled patches. By using \system, we can easily patch this bug at the source level and generate a patched recompiled binary.
This example highlights \system's value in bridging the gap between binary vulnerabilities and source-level patches, eliminating the need for error-prone manual assembly editing---particularly important when dealing with complex structures like nested loops.

\textbf{CVE-2020-16593.} \label{cve-2020-16593}
This null pointer dereference vulnerability~\cite{CVE-2020-16593} requires only a simple additional
check in the source code. However, the decompiler fails to recover the original local structs for
the variables {\tt var} and {\tt spec\_var} in Listing~\ref{lst:cve-2020-16593},
instead representing them as individual stack variables with pointer arithmetic.
Without \system's stack layout preservation, naive recompilation would produce incorrect stack offsets,
causing the patched binary to access wrong memory locations.
Despite this, \system successfully recompiles the 370-line function with complex stack structures,
ensuring pointer arithmetic accesses the correct offsets matching the original binary.

Listing~\ref{lst:cve-2020-16593} displays the decompiled patch. Since the real
structs ({\tt var} and {\tt spec\_var}) are not recovered by the decompiler,
patching requires an understanding of pointer arithmetic to locate and patch the faulty
instruction.  In our user study, all 18 participants successfully developed the
correct solution and patched the binary using \system. Notably, several
participants had no prior experience with decompiler-generated code,
highlighting \system's accessibility.

\begin{listing}[!ht]
\begin{minted}{diff}
-  free (tz0);
  ...
+  while (c = *p, c_isspace (c))
+    p++;
+  break;
\end{minted}
\begin{minted}{diff}
-  free(ptr);
  ...
+  char c;
+  while (c = *a2, c_isspace(c))
+    a2++;
+  break;
\end{minted}
    \caption{Source (top) and decompiled (bottom) patches for CVE-2014-9471}
    \label{lst:cve-2014-9471}
\end{listing}

\begin{listing}[!ht]
\begin{minted}{diff}
  if (var->name == NULL)
    var->name = spec_var->name;
- if (var->file == NULL)
+ if (var->file == NULL && spec_var->file != NULL)
    var->file = strdup (spec_var->file);
\end{minted}
\begin{minted}{diff}
  if ( !*((_QWORD *)v27 + 4) )
    *((_QWORD *)v27 + 4) = v37[4];
- if ( !*((_QWORD *)v27 + 2))
+ if ( !*((_QWORD *)v27 + 2) && v37[2] )
  {
    v15 = strdup((const char *)v37[2]);
    *((_QWORD *)v27 + 2) = v15;
\end{minted}
    \caption{Source (top) and decompiled (bottom) patches for CVE-2020-16593}
    \label{lst:cve-2020-16593}
\end{listing}

\section{Limitations and Future Work}
\textbf{Patch localization and translation.}
Compilation is inherently lossy and irreversible. While \system
corrects most syntactic and semantic issues in decompiled code, the result may
still differ structurally from the original source. Our evaluation of common vulnerability types showed that patch localization is practical: developers successfully applied patches within 10 minutes, and our LLM evaluation demonstrates this process can be fully automated. However, these results are based on vulnerabilities with well-defined patches. Future work could further improve patch localization for more complex scenarios, such as novel vulnerabilities or patches spanning multiple interdependent functions, which may remain challenging for both humans and LLMs.

\textbf{Better evaluation of correctness.}
To assess the correctness of decompiled code, we use output equivalence; that is,
for the same inputs, the recompiled code (after \system's fixes) should produce
outputs matching the original program. We evaluate this using available test
suites, though these may not fully cover all binary behaviors. For future work,
we propose two complementary directions: (1) using fuzzing to generate broader
test coverage, and (2) applying symbolic verification techniques, such as
VeriBin~\cite{veribin} or D-Helix~\cite{dhelix}, to provide stronger correctness
guarantees. However, integrating these tools would require substantial engineering
effort due to their legacy toolchain dependencies, which are incompatible with modern build
environments, and scalability constraints for large-scale evaluation.

\textbf{Binary size and performance impact.}
Recompiled binaries produced by \system are often larger than the original
program. The \system framework itself introduces minimal performance overhead:
only a trampoline jump when recompiled functions exceed their original size.
Since patching typically requires recompiling only a small set of functions, the
increase in size is generally minimal. Execution overhead depends on the specific
patch applied and approximates source-level instrumentation overhead. When
recompiling at higher optimization levels, performance may even improve. However,
size increases can still be relevant in resource-constrained systems, such as
embedded devices.

\textbf{Support for additional architectures.}
\system is architecture-agnostic by design: supporting
new architectures requires only minor configuration (calling conventions,
register naming) without algorithmic changes. However, our evaluation focused on
x86-64 due to mature toolchains and test infrastructure. Testing on ARM faced
challenges such as test-case instability and emulation overhead that hindered comprehensive
cross-architecture evaluation.

\section{Related Work}

\textbf{Binary rewriting.}
Binary rewriting modifies compiled binaries without source code access. Existing tools are often task-specific—e.g., for security hardening~\cite{programshepherding,rar,ppi,ccfir,cfci,patharmor,sbr,stir,gadgeme,oxymoron,secret,qiao_shadow}, fuzzing~\cite{angr,retrowrite,stochfuzz}, post-link optimization~\cite{bolt}, or debloating~\cite{bincfgtrim,razor}. In contrast, bug fixing requires a more flexible interface to patch custom code. Tools like PSI~\cite{psi}, \textsc{Egalito}~\cite{egalito}, \textsc{Safer}~\cite{safer}, e9Patch~\cite{e9patch}, and ARMore~\cite{armore} offer generic instrumentation APIs but operate at the assembly~\cite{dyninst,psi,armore,safer,e9patch} or IR~\cite{egalito} level, which is less human-friendly.

Dynamic instrumentation frameworks~\cite{pintool,dynamorio} support injecting C-level logic but incur high runtime overhead, making them unsuitable for patching. PRD~\cite{prd} introduced recompilation from decompiled code to enable source-level fault localization~\cite{rafl} and patch generation~\cite{apr-abhik,monperrus2018living,neural-repair,vulrepair}, but it does not address the semantic correctness of the recompiled output. \system targets this gap by improving the semantic quality of recompiled binaries.

\textbf{Evaluating Decompiler Output.}
Dramko et al.~\cite{dramko2024taxonomy} highlight readability challenges in modern decompilers, including incorrect return types and improper decomposition of structs—issues that \system already addresses to preserve semantic correctness. Liu et al.~\cite{howfarwevecome} analyze type, variable, and control-flow recovery issues but rely on manual inspection and synthetic datasets (e.g., from Csmith~\cite{csmith}) with simplified features, which limits the real-world applicability of their findings.
D-Helix~\cite{dhelix} introduces a symbolic testing framework to detect semantic errors by comparing symbolic expressions from binary IR and decompiled code. While it does not aim to fix issues or improve recompilation, it offers a foundation for error detection and future automated repair.

\textbf{Type and Data Structure Recovery.}
Recent works improve decompiler accuracy through machine learning and probabilistic methods for recovering types and data structures. Osprey~\cite{osprey} uses probabilistic analysis for variable and data structure recovery in stripped binaries. Resym~\cite{resym} and subsequent work~\cite{banerjee2021variable,xu2025unleashing} leverage large language models to recover variable names and data structure symbols from stripped binaries. While these approaches can improve initial decompilation quality, they focus on decompiler enhancement rather than addressing the recompilation challenges that \system targets. These methods are complementary to \system: improved type recovery can reduce decompilation errors, while \system's binary-aware recompilation ensures semantic correctness, even when type recovery is imperfect.

\textbf{Recompilation Approaches.}
DecLLM~\cite{decllm} uses large language models (LLMs) to improve decompilation
through an iterative repair loop guided by static and dynamic feedback. It
evaluates recompilation on Coreutils and DARPA CGC datasets, assuming access to
symbols and debug information. Under this setup, 58.2\% of failures are struct-related,
and DecLLM shows that LLMs can infer struct types when debug data is present. In
contrast, \system solves these issues more comprehensively (even in fully
stripped binaries) during the recompilation step by leveraging binary-aware
recompilation. DecLLM also struggles with long functions due
to context limitations, whereas \system recompiles functions over 2,000 lines
without loss in accuracy or performance.
The two approaches are complementary: \system uses binary-aware recompilation to
address common syntax and semantic issues, while DecLLM can fix rare syntax
errors outside \system's scope (Section~\ref{sec:challenges}). Combining both
could further increase the number of fully recompilable binaries.

Alternatively, Ramblr~\cite{ramblr} and ddisasm~\cite{ddisasm} produce reassemblable assembly through static and datalog-based analysis, respectively.
While they address pointer recovery and symbolic reference resolution, they still require assembly-level expertise to implement patches. In contrast, \system enables binary patching without requiring any assembly knowledge.

\section{Conclusion}

In this paper, we presented \system, a novel approach to enable practical static binary recompilation and patching.
By implementing a pipeline that addresses both syntactic and semantic issues in decompiled code, \system significantly improves the recompilability of decompiled functions.
We demonstrate \system's effectiveness by resolving approximately 81\% of previously incorrect functions produced by the Hex-Rays decompiler.
Furthermore, our case studies on real-world vulnerabilities, together with controlled studies involving both human participants and LLM systems, highlight \system{}'s effectiveness and practicality, with both humans and LLMs achieving 100\% patching success when using \system.

Overall, our results show that \system{} transforms binary patching from a task that is error-prone and requires specialized expertise into one that is practical and reliable for both humans and LLMs.
By bridging the gap between binary-level vulnerabilities and source-level patches, \system offers a powerful tool for maintaining and securing legacy or closed-source software systems, thereby opening new avenues for future research in this critical area of computer security.

\section*{Acknowledgment}
We would like to thank Prof.~Sergey Bratus and Prof.~Zion Leonahenahe Basque for their encouragement and technical feedback regarding this project.
This research was supported in part by ARPA-H under Award SP4701-23-C-0074, by the National Science Foundation under Grants 2442339, 2232915, and 2146568, by the Office of Naval Research under Grant N00014-23-1-2563, and by the Defense Advanced Research Projects Agency (DARPA) under contracts N66001-20-C-4031 and N66001-22-C-4026.
The U.S.\ Government is authorized to reproduce and distribute reprints for Governmental purposes notwithstanding any copyright notation thereon.
The views and conclusions contained herein are those of the authors and should not be interpreted as necessarily representing the official policies or endorsements, either expressed or implied, of ARPA-H, DARPA, NSF, ONR, or the U.S.\ Government.

\bibliographystyle{IEEEtran}
\bibliography{references}

\appendices
\section{Evaluated CVE Details}
\label{appendix:cve_details}

This appendix provides detailed information about the real-world CVEs used to evaluate \system's patching capabilities, including their classification, affected programs, and patching results.

\begin{table}[H]
\centering
\caption{Real-world CVEs evaluated with \system. LOC indicates lines of code in the patched function. CWE(s) indicates Common Weakness Enumeration classification(s). Res. indicates result (Y=success, N=failure).}
\label{tab:cve_details}
\scriptsize
\begin{tabular}{|l|l|l|l|r|c|}
\hline
\textbf{CVE-ID} & \textbf{Dataset} & \textbf{Program} & \textbf{CWE(s)} & \textbf{LOC} & \textbf{Res.} \\
\hline
CVE-2014-9471 & coreutils & date & N/A & \num{336} & Y \\
CVE-2024-0684 & coreutils & split & 787, 122 & \num{139} & Y \\
CVE-2020-16590 & binutils & readelf & 415 & \num{274} & Y \\
CVE-2020-16591 & binutils & readelf & 125 & \num{274} & Y \\
CVE-2020-16592 & binutils & nm & 416 & \num{118} & Y \\
CVE-2020-16593 & binutils & addr2line & 476 & \num{370} & Y \\
CVE-2020-16599 & binutils & nm & 476 & \num{64} & Y \\
CVE-2021-20284 & binutils & nm & 787, 119 & \num{243} & N \\
CVE-2021-20294 & binutils & readelf & 787 & \num{140} & Y \\
CVE-2022-35205 & binutils & readelf & 617 & \num{836} & Y \\
CVE-2022-35206 & binutils & readelf & 476 & \num{2145} & Y \\
CVE-2022-4285 & binutils & nm & 476 & \num{282} & Y \\
CVE-2022-48063 & binutils & objdump & 400 & \num{112} & Y \\
CVE-2023-1972 & binutils & objdump & 787, 119 & \num{286} & Y \\
\hline
\end{tabular}
\end{table}

\section{Screening Questionnaire}
\label{appendix:screening}

This appendix contains the screening questionnaire used to assess participant backgrounds and experience levels before the user study.

{\footnotesize
\begin{enumerate}
    \item Are you 18 years of age or older?
    \begin{itemize}
        \item Yes
        \item No
    \end{itemize}

    \item What is your level of experience with Linux/Unix and shell scripting?
    \begin{itemize}
        \item No experience
        \item Basic (familiar with common commands like cat, ls, cd, etc. and comfortable with bash scripts)
        \item Advanced (extensive system administration and scripting)
    \end{itemize}

    \item What is your level of experience with C programming?
    \begin{itemize}
        \item No experience
        \item Basic (can write simple C programs)
        \item Advanced (extensive experience with complex applications)
    \end{itemize}

    \item What is your experience with common memory-related vulnerabilities in C?
    \begin{itemize}
        \item No experience
        \item Basic (understand concepts like buffer overflow, use-after-free)
        \item Advanced (experience exploiting memory vulnerabilities)
    \end{itemize}

    \item What is your experience with decompiler yrrxpv?
    \begin{itemize}
        \item No experience
        \item Basic (this decompiler doesn't exist, you should select no experience)
        \item Advanced (same as basic, but have more than 1 year of experience)
    \end{itemize}

    \item Which other decompilers have you used? (Select all that apply)
    \begin{itemize}
        \item Ghidra
        \item IDA Pro
        \item Binary Ninja
        \item Others (Please specify): \underline{\hspace{4cm}}
    \end{itemize}

    \item How comfortable are you with using GNU Debugger (GDB)?
    \begin{itemize}
        \item No experience
        \item Basic (familiar with common debugging commands)
        \item Advanced (extensive debugging experience)
    \end{itemize}

    \item Are you able to commit to a 3.5 hour study session on Zoom? You'll need a computer with a stable internet connection and an up-to-date Chrome browser.
    \begin{itemize}
        \item Yes
        \item No
    \end{itemize}

    \item Which declaration creates a pointer to an integer?
    \begin{itemize}
        \item \texttt{ptr* int;}
        \item \texttt{int* ptr;}
        \item \texttt{pointer int;}
    \end{itemize}

    \item What does the arrow operator (->) do in C?
    \begin{itemize}
        \item Access structure members through a pointer
        \item Moves data from one pointer to another
        \item Compares two pointers
    \end{itemize}

    \item How do you convert (cast) one type to another in C?
    \begin{itemize}
        \item (int) x
        \item int(x)
        \item convert.int(x)
    \end{itemize}

    \item In this snippet from git diff, which line was actually removed from the original code?
    \begin{minted}{diff}
    void process() {
   -    log("starting");
        int count = 0;
   +    log("begin processing");
        return;
    }
    \end{minted}
    \begin{itemize}
        \item \texttt{void process() \{}
        \item \texttt{log("starting");}
        \item \texttt{int count = 0;}
        \item \texttt{log("begin processing");}
    \end{itemize}


\end{enumerate}
}

\raggedbottom
\section{User Study Completion Time}
\label{appendix:study_time}

Note that the times were recorded manually with a stopwatch, which may be subject to up to 5 seconds of error. The time is recorded in minutes:seconds format.
\begin{table}[H]
    \centering
    \caption{Time taken for each participant to locate, patch, and fix vulnerability 1 with and without \system}
    \footnotesize
    \begin{tabular}{|c|c|c|c|c|c|c|}
    \hline
        \multirow{2}{*}{} & \multicolumn{3}{c|}{With \system} & \multicolumn{3}{c|}{W/O \system} \\
        \cline{2-7}
        ~ & Locate & Patch & Fix & Locate & Patch & Fix \\ \hline
        \multirow{18}{*}{Time} & 0:18 & 0:25 & 0:42 & 0:14 & 0:14 & Failed \\ \cline{2-7}
        ~ & 0:05 & 0:12 & 0:20 & 0:31 & 0:55 & Failed \\ \cline{2-7}
        ~ & 2:48 & 3:40 & 4:31 & 0:21 & 0:21 & Failed \\ \cline{2-7}
        ~ & 0:12 & 0:23 & 1:57 & 4:34 & 7:22 & Failed \\ \cline{2-7}
        ~ & 2:45 & 2:48 & 3:10 & 8:45 & 8:56 & Failed \\ \cline{2-7}
        ~ & 0:07 & 0:15 & 0:44 & 0:37 & 0:51 & Failed \\ \cline{2-7}
        ~ & 1:42 & 2:15 & 2:56 & 0:10 & 0:10 & Failed \\ \cline{2-7}
        ~ & 3:32 & 4:18 & 5:04 & 0:06 & 0:48 & Failed \\ \cline{2-7}
        ~ & 0:45 & 2:23 & 2:55 & 1:46 & 2:56 & Failed \\ \cline{2-7}
        ~ & 1:38 & 3:12 & 3:28 & 0:11 & 0:17 & Failed \\ \cline{2-7}
        ~ & 0:18 & 0:44 & 1:22 & 5:03 & 11:43 & Failed \\ \cline{2-7}
        ~ & 1:32 & 1:54 & 2:12 & 0:13 & 0:22 & Failed \\ \cline{2-7}
        ~ & 0:38 & 1:12 & 2:18 & 0:24 & 0:35 & 15:44 \\ \cline{2-7}
        ~ & 0:09 & 0:18 & 0:33 & 2:24 & 3:34 & Failed \\ \cline{2-7}
        ~ & 0:18 & 0:26 & 0:48 & 1:30 & 2:09 & 17:38 \\ \cline{2-7}
        ~ & 0:12 & 0:18 & 0:42 & 1:12 & 1:42 & Failed \\ \cline{2-7}
        ~ & 1:33 & 2:09 & 2:47 & 0:12 & 0:21 & Failed \\ \cline{2-7}
        ~ & 0:40 & 0:47 & 1:22 & 0:15 & 0:27 & Failed \\ \hline
        Avg & 1:04 & 1:32 & 2:06 & 1:34 & 2:25 & 16:41 \\ \hline
    \end{tabular}
\end{table}

\begin{table}[H]
    \centering
    \caption{Time taken for each participant to locate, patch, and fix vulnerability 2 with and without \system}
    \footnotesize
    \begin{tabular}{|c|c|c|c|c|c|c|}
    \hline
        \multirow{2}{*}{} & \multicolumn{3}{c|}{With \system} & \multicolumn{3}{c|}{W/O \system} \\
        \cline{2-7}
        ~ & Locate & Patch & Fix & Locate & Patch & Fix \\ \hline
        \multirow{18}{*}{Time} & 0:27 & 0:35 & 0:57 & 0:12 & 0:12 & Failed \\ \cline{2-7}
        ~ & 0:10 & 0:16 & 0:24 & 1:57 & 2:22 & Failed \\ \cline{2-7}
        ~ & 0:37 & 0:51 & 1:32 & 0:17 & 0:17 & Failed \\ \cline{2-7}
        ~ & 0:07 & 0:14 & 0:52 & 0:34 & 0:45 & Failed \\ \cline{2-7}
        ~ & 0:06 & 0:09 & 0:20 & 1:32 & 2:06 & Failed \\ \cline{2-7}
        ~ & 0:07 & 0:15 & 1:01 & 0:19 & 2:01 & Failed \\ \cline{2-7}
        ~ & 1:04 & 1:24 & 2:17 & 0:11 & 0:22 & Failed \\ \cline{2-7}
        ~ & 0:53 & 1:20 & 2:14 & 0:07 & 0:22 & Failed \\ \cline{2-7}
        ~ & 3:30 & 3:48 & 4:47 & 5:07 & 6:04 & Failed \\ \cline{2-7}
        ~ & 0:17 & 0:54 & 1:44 & 0:09 & 0:14 & Failed \\ \cline{2-7}
        ~ & 0:12 & 0:31 & 1:31 & 7:23 & 11:22 & Failed \\ \cline{2-7}
        ~ & 0:22 & 1:46 & 2:52 & 0:08 & 0:19 & Failed \\ \cline{2-7}
        ~ & 1:06 & 2:20 & 5:43 & 0:09 & 0:15 & Failed \\ \cline{2-7}
        ~ & 0:08 & 0:14 & 0:50 & 1:42 & 2:07 & Failed \\ \cline{2-7}
        ~ & 0:13 & 0:16 & 0:48 & 0:47 & 1:07 & Failed \\ \cline{2-7}
        ~ & 0:08 & 0:16 & 0:54 & 0:24 & 0:42 & Failed \\ \cline{2-7}
        ~ & 1:20 & 1:42 & 3:44 & 0:09 & 0:22 & Failed \\ \cline{2-7}
        ~ & 0:32 & 0:52 & 1:41 & 3:20 & 3:52 & Failed \\ \hline
        Avg & 0:37 & 0:59 & 1:53 & 1:21 & 1:56 & N/A \\ \hline
    \end{tabular}
\end{table}

\begin{table}[H]
    \centering
    \caption{Time taken for each participant to locate, patch, and fix vulnerability 3 with and without \system}
    \footnotesize
    \begin{tabular}{|c|c|c|c|c|c|c|}
    \hline
        \multirow{2}{*}{} & \multicolumn{3}{c|}{With \system} & \multicolumn{3}{c|}{W/O \system} \\
        \cline{2-7}
        ~ & Locate & Patch & Fix & Locate & Patch & Fix \\ \hline
        \multirow{18}{*}{Time} & 6:06 & 6:47 & 7:25 & 0:19 & 0:19 & Failed \\ \cline{2-7}
        ~ & 0:11 & 0:24 & 0:53 & 1:02 & 3:15 & Failed \\ \cline{2-7}
        ~ & 6:20 & 9:32 & 10:07 & 0:23 & 0:23 & Failed \\ \cline{2-7}
        ~ & 0:24 & 0:24 & 6:13 & 1:55 & 4:58 & Failed \\ \cline{2-7}
        ~ & 0:07 & 0:24 & 1:26 & 8:23 & 10:24 & Failed \\ \cline{2-7}
        ~ & 0:12 & 0:25 & 1:01 & 5:11 & 8:06 & Failed \\ \cline{2-7}
        ~ & 3:03 & 6:18 & 13:10 & 0:11 & 0:19 & Failed \\ \cline{2-7}
        ~ & 1:06 & 2:48 & 5:44 & 1:50 & 2:11 & Failed \\ \cline{2-7}
        ~ & 0:44 & 1:31 & 2:10 & 11:45 & 18:45 & Failed \\ \cline{2-7}
        ~ & 0:12 & 4:21 & 5:53 & 0:08 & 0:13 & Failed \\ \cline{2-7}
        ~ & 0:12 & 1:23 & 3:20 & 7:53 & Failed & Failed \\ \cline{2-7}
        ~ & 1:47 & 8:12 & 9:17 & 1:21 & 1:36 & Failed \\ \cline{2-7}
        ~ & 2:28 & 6:40 & 8:19 & 0:31 & 0:46 & Failed \\ \cline{2-7}
        ~ & 0:17 & 0:33 & 1:07 & 1:52 & 10:32 & Failed \\ \cline{2-7}
        ~ & 0:20 & 0:58 & 6:44 & 1:47 & 3:20 & Failed \\ \cline{2-7}
        ~ & 0:14 & 0:22 & 1:21 & 10:44 & 12:09 & Failed \\ \cline{2-7}
        ~ & 1:12 & 3:28 & 7:53 & 0:07 & 0:32 & Failed \\ \cline{2-7}
        ~ & 0:17 & 0:27 & 1:24 & 1:24 & 5:10 & Failed \\ \hline
        Avg & 1:24 & 3:03 & 5:11 & 3:09 & 4:52 & N/A \\ \hline
    \end{tabular}
\end{table}
\flushbottom

\newpage
\section{LLM Evaluation Results}
\label{appendix:llm_results}

Table~\ref{tab:llm_results} shows detailed LLM evaluation results. Each model was evaluated with 3 trials per vulnerability per condition (with/without \system), totaling 27 trials.

\begin{table}[H]
    \centering
    \caption{LLM evaluation results showing average completion times and patch attempts per vulnerability. With \system, all models achieved 100\% success (27/27 trials) averaging 23.5 seconds and 1.0 patch attempt. Without \system, all models achieved 0\% success (0/27 trials), averaging 557.0 seconds and 13.9 patch attempts before timeout.}
    \label{tab:llm_results}
    \scriptsize
    \setlength{\tabcolsep}{4pt}
    \begin{tabular}{|l|c|c|c|c|c|}
    \hline
        \textbf{Model} & \textbf{Vuln} & \multicolumn{2}{c|}{\textbf{With \system}} & \multicolumn{2}{c|}{\textbf{Without \system}} \\
        \cline{3-6}
        ~ & ~ & \textbf{Time(s)} & \textbf{Attempts} & \textbf{Time(s)} & \textbf{Attempts} \\
        \hline
        \multirow{4}{*}{Claude 4.5}
        & Easy   & 13.9 & 1.0 & 160.7 & 7.0 \\
        & Medium & 10.1 & 1.0 & 125.1 & 1.7 \\
        & Hard   & 13.1 & 1.0 & 214.0 & 9.3 \\
        \cline{2-6}
        & \textbf{Overall} & \textbf{12.4} & \textbf{1.0} & \textbf{166.6} & \textbf{6.0} \\
        \hline
        \multirow{4}{*}{Gemini 2.5}
        & Easy   & 34.5 & 1.0 & 429.0 & 7.3 \\
        & Medium &  7.1 & 1.0 & 438.6 & 5.7 \\
        & Hard   & 12.6 & 1.0 & 274.8 & 4.3 \\
        \cline{2-6}
        & \textbf{Overall} & \textbf{18.0} & \textbf{1.0} & \textbf{380.8} & \textbf{5.8} \\
        \hline
        \multirow{4}{*}{GPT-5}
        & Easy   & 36.4 & 1.0 & 1384.7 & 30.0 \\
        & Medium & 39.8 & 1.3 &  716.3 & 30.0 \\
        & Hard   & 44.3 & 1.0 & 1270.0 & 30.0 \\
        \cline{2-6}
        & \textbf{Overall} & \textbf{40.2} & \textbf{1.1} & \textbf{1123.7} & \textbf{30.0} \\
        \hline
        \hline
        \multicolumn{2}{|c|}{\textbf{All Models}} & \textbf{23.5} & \textbf{1.0} & \textbf{557.0} & \textbf{13.9} \\
        \hline
    \end{tabular}
\end{table}

\end{document}